\documentclass{article}
\usepackage{cite}
\usepackage{authblk}
\usepackage{graphicx}
\usepackage[a4paper, left=2.5cm, right=2.5cm, top=3cm, bottom=3cm]{geometry}
\usepackage{dcolumn}
\usepackage{bm}

\usepackage[utf8]{inputenc}
\usepackage[T1]{fontenc}
\usepackage{mathptmx}
\usepackage{etoolbox}
\usepackage{multirow}
\usepackage{booktabs}
\usepackage{amsmath}

\begin{document}

\title{Machine Learning Informed by Micro and Mesoscopic Statistical Physics Methods for Community Detection} 

\author[1]{Yijun Ran\thanks{Yijun Ran and Junfan Yi contributed equally to this work.}}

\author[2]{Junfan Yi\thanks{Yijun Ran and Junfan Yi contributed equally to this work.}}%

\author[2]{Wei Si}

\author[3,4]{Michael Small\thanks{Email: \texttt{michael.small@uwa.edu.au}}}

\author[2]{Ke-ke Shang\thanks{Email: \texttt{kekeshang@nju.edu.cn}, \texttt{keke.shang.1989@gmail.com}}}

\affil[1]{Center for Computational Communication Research, Beijing Normal University, Zhuhai, 519087, P. R. China}
\affil[2]{Computational Communication Collaboratory, Nanjing University, Nanjing, 210093, P.R. China}%
\affil[3]{Complex Systems Group, Department of Mathematics and Statistics, The University of Western Australia, Crawley, Western Australia 6009, Australia}
\affil[4]{Mineral Resources, CSIRO, Kensington, WA, 6151, Australia}%

\date{}

\maketitle

\begin{abstract}
Community detection plays a crucial role in understanding the structural organization of complex networks. Previous methods, particularly those from statistical physics, primarily focus on the analysis of mesoscopic network structures and often struggle to integrate fine-grained node similarities. To address this limitation, we propose a low-complexity framework that integrates machine learning to embed micro-level node-pair similarities into mesoscopic community structures. By leveraging ensemble learning models, our approach enhances both structural coherence and detection accuracy. Experimental evaluations on artificial and real-world networks demonstrate that our framework consistently outperforms conventional methods, achieving higher modularity and improved accuracy in NMI and ARI. Notably, when ground-truth labels are available, our approach yields the most accurate detection results, effectively recovering real-world community structures while minimizing misclassifications. To further explain our framework's performance, we analyze the correlation between node-pair similarity and evaluation metrics. The results reveal a strong and statistically significant correlation, underscoring the critical role of node-pair similarity in enhancing detection accuracy. Overall, our findings highlight the synergy between machine learning and statistical physics, demonstrating how machine learning techniques can enhance network analysis and uncover complex structural patterns.

\end{abstract}


\section{Introduction}
\label{sec:level1}

Community detection, a central problem in network science, plays a crucial role in understanding the underlying structures and relationships within complex systems \cite{fortunato202220,kojaku2024network,zhang2014scalable,hoffmann2020community,peixoto2019network}. Networks ranging from social media platforms to biological systems often exhibit modular structures where nodes within a community are more densely connected to others than to nodes outside the community \cite{ahn2010link,ghasemian2016detectability,grilli2016modularity}. These modular formations or communities often reflect functional or structural groupings that can provide insights into the dynamics of the system \cite{lambiotte2021modularity,zhou2006hierarchical,red2011comparing}. However, detecting these communities efficiently and accurately remains a challenging task, especially in large-scale or sparse networks, where traditional methods may fail to capture intricate patterns of connectivity \cite{su2022comprehensive,jin2021survey,van2019scalable}. 

Traditional community detection approaches often rely on optimizing a global objective function, such as modularity or conductance, which evaluates the quality of a community partition based on the overall link density within communities compared to between communities \cite{newman2004finding,ghasemian2019evaluating,xie2013overlapping}. While these methods have proven successful in many contexts, they are primarily topological in nature and do not fully leverage the wealth of information that can be extracted from the relationships between individual nodes \cite{de2017community,shang2020novel}. In particular, node-pair similarities, which capture subtle and often overlooked relationships between nodes, remain largely unexplored in community detection.

Recent advancements suggest that incorporating micro-level information, such as node-pair similarities, can provide deeper insights into the structure of a network \cite{valles2016multilayer,ghasemian2020stacking}. These similarities, which reflect the extent to which two nodes exhibit analogous behavior or share common attributes, can significantly enhance the understanding of community formation \cite{guimera2020one,shang2022link}. For example, nodes within the same community are likely to exhibit similar interaction patterns or share similar features, and these similarities can serve as valuable cues for detecting community boundaries. However, directly integrating these microscopic similarities into the community detection process presents several challenges, as the underlying community structure is inherently a mesoscopic phenomenon involving interdependencies across many nodes.

To address this gap, we propose a novel framework that integrates machine learning techniques to seamlessly embed microscopic information---specifically, the similarities between node pairs---into the mesoscopic community structure. We utilize advanced ensemble learning models to learn a mapping that combines node pairs with community-level structural features. By capturing these similarities, we enrich the representation of each node in the network, enabling a more nuanced understanding of the inter-node relationships. This enriched representation is then used to guide community detection, improving the accuracy and robustness of the detected communities. 

By leveraging ensemble learning models, our framework refines the representation of network structures, leading to more accurate and robust community detection. This approach not only enhances modularity but also improves the alignment between detected communities and ground-truth labels, as demonstrated by extensive experiments on artificial and real-world networks. The integration of machine learning with statistical physics provides a more comprehensive understanding of network organization, bridging the gap between local interactions and global structures. 

\section{Preliminaries}
\label{sec:level2}

\subsection{Problem definition}
\label{section21}
The problem of community detection in complex networks involves identifying groups of nodes, known as communities, within a network \cite{liu2022dynamic,xiao2024higher,wang2023non}. In this context, we consider an undirected simple network $G=(N, L)$, where $N$ represents the set of nodes and $L$ represents the set of links connecting pairs of nodes. A community is defined as a subset of nodes $C=\{n_1, n_2,..., n_k\}$, where the nodes within the community are more densely connected to others than to nodes outside the community.

The goal of community detection is to partition the entire set of nodes $N$ into disjoint communities, denoted as $C_1, C_2,..., C_k$, such that the union of these communities covers all nodes in the network \cite{pinto2024optimizing,xiao2024constrained}. In this study, we focus on detecting crisp communities, where each node is assigned to exactly one community. This approach assumes that the network structure is best understood through exclusive, non-overlapping groups.

\subsection{Detection methods}
\label{section22}
We consider four classic community detection methods to validate our hypothesis that the similarities between node pairs can provide critical information to improve the accuracy of the detected communities \cite{fortunato2010community,fortunato2016community}. The Louvain algorithm is a widely used community detection method that aims to optimize modularity, a measure of the quality of a network partition \cite{traag2019louvain}. It operates in two phases: the first involves local optimization, where each node is initially assigned to its community, and then iteratively moved to the community that increases the overall modularity. The second phase aggregates the nodes into super-nodes, with each super-node representing a community, and the process is applied recursively to detect communities at different levels of the network. While Louvain is efficient and capable of handling large networks, it can suffer from the resolution limit, making it less effective at detecting small communities within larger ones.

The Leiden algorithm is an enhanced version of the Louvain algorithm designed to address some of its shortcomings \cite{traag2019louvain}. One major improvement is its ability to ensure that all communities remain connected, even after optimization, thus preventing the creation of disconnected communities, which can happen in Louvain. Additionally, the Leiden algorithm introduces a refinement step that improves the stability of community detection and reduces the likelihood of producing poor or suboptimal partitions. This makes the Leiden algorithm faster, more reliable, and more accurate than Louvain, particularly when dealing with large, complex networks. The Leiden algorithm is considered to provide higher-quality community partitions and greater computational efficiency, especially in large-scale networks with complex structures.

The Fast-Greedy (FG) algorithm is another modularity-based method that greedily optimizes the network division \cite{clauset2004finding,raj2020information}. It starts with each node in its community and calculates the modularity for all possible mergers of communities. The two communities that result in the greatest increase in modularity are then merged, and this process is repeated until no further merges yield an increase in modularity. While the Fast-Greedy algorithm is computationally efficient and effective for large networks, it may still miss smaller communities and can be prone to producing less stable solutions when compared to other methods.

The Infomap algorithm takes a different approach, relying on information theory to detect community structures \cite{rosvall2008maps}. It minimizes the description length of a random walk over the network by using the map equation, which quantifies the amount of information needed to describe the movement of a random walker. The algorithm aims to localize the random walker within communities, thus minimizing the information transmitted across community boundaries. The Infomap is particularly well-suited for detecting hierarchical and overlapping community structures and is less affected by the resolution limit than modularity-based methods like Louvain and Fast-Greedy. It is considered one of the most effective methods for uncovering intricate community structures in large networks.

\subsection{Evaluation metrics}
\label{section23}
To evaluate the efficiency of our proposed framework, we use three commonly employed metrics in community detection: modularity \cite{newman2004finding,newman2006modularity,shi2024local}, normalized mutual information (NMI) \cite{zhang2015evaluating,zhu2021unsupervised}, and adjusted rand index (ARI) \cite{zhou2021robustecd,zhao2024detecting}. Modularity $Q$ is a widely used metric for evaluating the quality of community structure in a network. It quantifies the difference between the observed and expected number of links within communities, aiming to identify partitions where internal connectivity is significantly stronger than expected by chance. Modularity is particularly useful for networks without predefined labels and is defined as
\begin{equation}
Q = \sum_{i=1}^k \left[\frac{|L_{i}^{in}|}{|L|} - \left(\frac{d_i}{2|L|}\right)^2 \right],
\label{equation:q}
\end{equation}
where $k$ is the number of communities, $L_{i}^{in}$ is the sets of internal links within the community $i$. The total degree of all nodes in community $i$, denoted as $d_i$ is given by $d_i=2|L_{i}^{in}|+|L_{i}^{out}|$, where $L_{i}^{out}$ is the sets of external links connecting community $i$ to other communities. For a weighted similarity network, modularity can be naturally extended by replacing link counts with their respective weights. The weighted modularity is given by 
\begin{equation}
Q^w = \sum_{i=1}^k \left[\frac{W_i^{in}}{W} - \left(\frac{W_i}{2W}\right)^2 \right],
\label{equation:qw}
\end{equation}
where $W$ denotes the total sum of link weights in a weighted similarity network, $W_{i}^{in}$ represents the sum of the weights of internal links within community $i$. The total weight of community $i$, denoted as $W_i$ is given by $W_i=2W_{i}^{in}+W_{i}^{out}$, where $W_{i}^{out}$ is the sum of the weights of external links connecting community $i$ to other communities. Eq. \ref{equation:q} is a special case of Eq. \ref{equation:qw} when the weight of each link is 1.

A higher modularity value, approaching 1, indicates a stronger community structure, with a significantly higher density of internal links compared to random partitions. This suggests that the nodes within a community are more tightly connected than would be expected in a random configuration.

The NMI is a metric used to evaluate the efficiency of community detection algorithms, particularly when the ground-truth division of a network is available. It quantifies the amount of information shared between two divisions---one from the community detection algorithm and the other from the real, ground-truth division. The value of NMI ranges from 0 to 1, with higher values indicating better alignment between the detected community structure and the actual ground-truth structure. Essentially, NMI measures the similarity between the experimental community division and the real division, with a higher NMI suggesting that the algorithm has performed better in identifying the true communities within the network. NMI accounts for the inherent size and structure of the network, making it a normalized measure, thus allowing for more consistent comparisons across different networks and divisions.

The ARI is another commonly used metric for evaluating community detection methods. It corrects the chance of random assignments by adjusting for the expected similarity between random partitions, thus providing a more robust evaluation of the community detection algorithm's performance. The ARI ranges from -1 to 1, where a value close to 1 indicates perfect agreement between the algorithm's detected communities and the ground-truth communities, while a value close to 0 suggests that the algorithm's community assignments are no better than random. Negative ARI values indicate that the algorithm's detected community structure is worse than random. Higher ARI values indicate more accurate community division, reflecting a higher percentage of correctly clustered nodes.

\section{Methodology}
\label{sec:level3}
\begin{figure}
\centering
\includegraphics[width=0.5\textwidth]{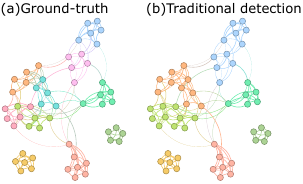}
\caption{A real-world network illustrating the limitations of existing community detection algorithms. Nodes in different colors denote distinct communities.
(a) The ground-truth community structure of the LOL network (with more details in Table~\ref{tab:table1}).
(b) The visualization of the LOL network’s communities detected via the Leiden algorithm. }
\label{fig1}
\end{figure}

Community detection has evolved significantly over the past two decades, leading to the development of numerous robust algorithms. Despite these advancements, challenges remain particularly in accurately classifying nodes within real-world networks. These misclassifications result in deviations that can impact practical applications. For example, as illustrated in Fig.~\ref{fig1}, classic detection algorithms may incorrectly assign cyan nodes (actual members of one community in Fig.~\ref{fig1}(a)) as yellow, or mislabel pink nodes as blue. These errors highlight the ongoing difficulties in achieving precise community detection in complex networks. This underscores the need for continued research to enhance the accuracy of these algorithms.

\begin{figure}
\centering
\includegraphics[width=0.5\textwidth]{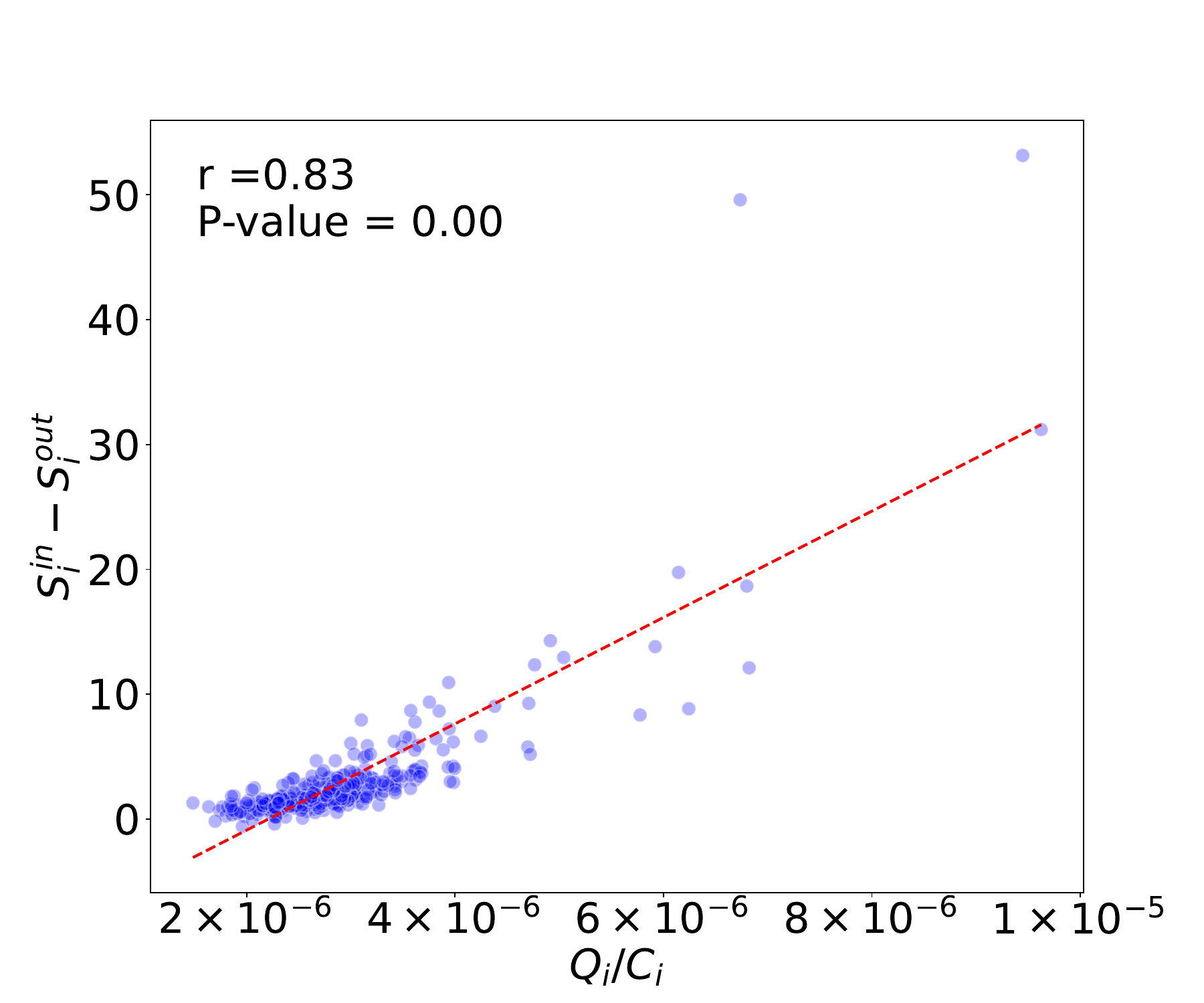}
\caption{The node-pair similarity related to network modularity. The network modularity $Q$ shows a strong correlation with node-pair similarity, as empirically confirmed. $S_i^{in}$ is normalized by the sum of similarities between all node pairs and the total number of internal links within the community $i$. Similarly, $S_i^{out}$ is normalized by the sum of similarities between all node pairs and the total number of external links connecting the community $i$ to others. 
$Q_i$ denotes the modularity of the community $i$, and $C_i$ represents the number of nodes within the community $i$. To validate this assumption, we use the DBLP network (described in Table~\ref{tab:table1}), with communities detected using the Infomap algorithm. The r is the Pearson correlation coefficient, and the P-value is from the Student's t test.}
\label{fig:sim}
\end{figure}

In this study, we propose that node-pair similarity is a valuable factor that can improve community detection performance. Recent research indicates that micro-level information can provide deeper insights into network structures \cite{valles2016multilayer,ghasemian2020stacking}. To assess the role of node-pair similarity, we examine its relationship with modularity $Q$ using real-world networks. Higher modularity values reflect a more meaningful network partitioning, where nodes within the same community are more densely connected, while nodes between communities are less connected. Hence, we calculate the difference between the average similarity (the value is quantified by Eq. \ref{equation:nn}) of node pairs within a community and that of pairs across communities, and then analyze how this difference correlates with modularity. Our findings show a strong positive correlation---the higher the modularity of a community, the greater the difference in similarity between internal and external node pairs (Fig. \ref{fig:sim}). This suggests that node-pair similarity is an important factor in enhancing community detection performance.

\begin{figure*}[htbp]
\centering
\includegraphics[width=\textwidth]{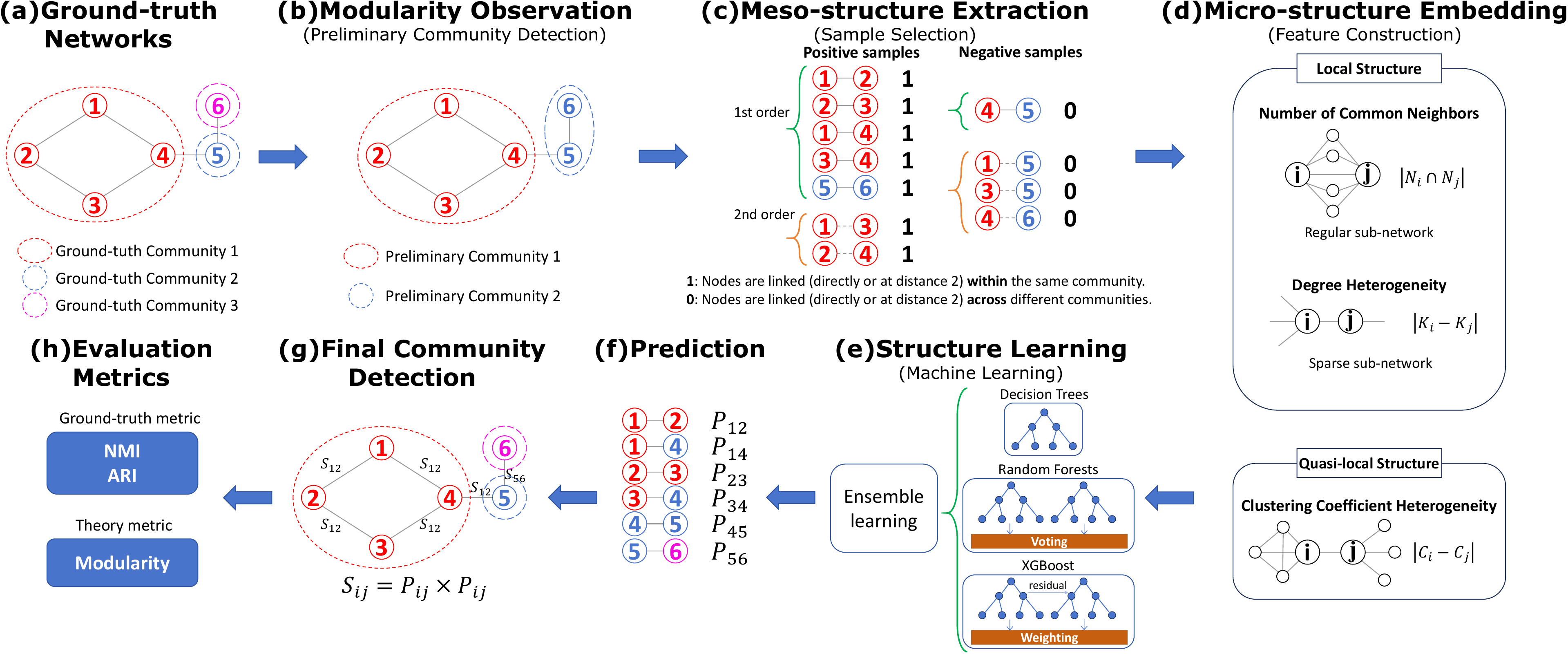}
\caption{Overview of the proposed detection framework. In the absence of ground-truth communities, we implement the community detection framework using a statistical-physics approach. (b) We apply four commonly used community detection algorithms (section \ref{section22}) to identify communities within the network, treating these detected communities as the ground truth. (c) We extract mesoscopic structural information from detected communities. Specifically, we sample first-order and second-order node pairs, classifying intra-community pairs as one category and inter-community pairs as another. These samples form the training and testing datasets for machine learning. (d) We introduce three microscopic structural features tailored to different real-world networks. (e) Classical machine learning methods are employed to predict node-pair similarity, while ensemble learning is used to enhance the performance of individual predictors. (f) A five-fold cross-validation strategy is applied to estimate similarity values for each node pair. (g) The predicted similarity values are integrated into the original network to facilitate community detection. (h) Finally, we evaluate the effectiveness of the proposed framework using three widely adopted performance metrics. Note that when ground-truth community information is available, step (b) can be omitted. We refer to this approach as the ground-truth method.}
\label{fig2}
\end{figure*}

To efficiently leverage the similarity between node pairs, we propose a novel framework that integrates machine learning techniques to incorporate microscopic information---specifically, the similarities between node pairs---into the mesoscopic community structure (Fig. \ref{fig2}). Our framework relies on the availability of ground-truth community data. In the absence of such information, we employ a statistical-physics detection algorithm to infer the communities, as illustrated in Fig. \ref{fig2}(b). This approach offers a mesoscopic perspective on community detection. Statistical physics provides a robust framework for analyzing complex systems. By applying principles from statistical mechanics, we can model the community detection process as a physical system, enabling the identification of community structures through methods such as Infomap and optimization of modularity. 

{\bf Mesoscopic structural information extraction}. Once the community structure is determined, we treat node pairs within the same community as one class and those in different communities as another, forming a binary classification problem that captures essential mesoscopic structural information. To quantify the similarity between node pairs, we propose an efficient sampling method that generates training and testing datasets by selecting both first-order and second-order node pairs (Fig. \ref{fig2}(c)).

{\bf Microscopic structure embedding for Sparse and Dense Networks}. Our framework accommodates the different structural properties of sparse and dense networks, which are commonly encountered in real-world systems. We consider three types of microscopic structural features to characterize these networks (Fig. \ref{fig2}(d)).
\begin{enumerate}
	\item Degree heterogeneity in sparse networks: In sparse networks, degree heterogeneity serves as a key feature \cite{shang2019link,shang2022link}. Specifically, a larger degree difference between two nodes suggests a higher likelihood that they belong to the same community. This heterogeneity is defined as:
    \begin{equation}
    R_{ij} = \lvert {K_i - K_j} \rvert, 
    \label{equation:kk}
    \end{equation}
    where $K_i$ is the degree of node $i$.
	
	\item Clustering coefficient heterogeneity: In addition to degree heterogeneity, we also examine the heterogeneity in clustering coefficients \cite{shang2019link,shang2022link}, which indicates that a greater difference in clustering coefficients between two nodes increases the likelihood that they belong to the same community. The clustering-coefficient heterogeneity is defined as:
    \begin{equation}
    R_{ij} = \lvert {C_i - C_j} \rvert, 
    \label{equation:cc}
    \end{equation}
    where $C_i=\frac{2T_i}{K_i(K_i-1)}$ is the clustering coefficient of node $i$. $T_i$ represents the number of triangles passing through node $i$. 
	
	\item Common neighbors in dense networks: For dense networks, we rely on the well-known common neighbor feature \cite{ran2024maximum}, which posits that two nodes sharing more neighbors are more likely to belong to the same community. This is defined as:
    \begin{equation}
     R_{ij} = \lvert {N_i \cap N_j} \rvert, 
     \label{equation:nn}
    \end{equation}
    where $N_i$ denotes the set over all neighbors of node $i$.
\end{enumerate}

{\bf Machine Learning for Community Detection}. To effectively combine mesoscopic and microscopic structural information, we employ machine learning algorithms to estimate the likelihood that a node pair belongs to the same community (Fig.~\ref{fig2}(e)). Specifically, we utilize three popular and efficient algorithms:
\begin{enumerate}
    \item Decision Trees (DT) \cite{zhang2022interpretable}: Decision trees are widely used for classification and regression tasks. They create a tree-like structure, where each internal node represents a decision point based on a specific feature, leading to a leaf node that signifies the outcome. Although decision trees are interpretable, they tend to overfit, which can be mitigated by pruning.

    \item Random Forests (RF) \cite{thakur2024machine}: Random forests improve upon decision trees by constructing an ensemble of trees, each trained on random subsets of data and features using the bagging method. This ensemble approach reduces overfitting and enhances generalization, offering robust performance, even in high-dimensional spaces.

    \item XGBoost (XGB) \cite{chen2016xgboost}: XGBoost is an advanced gradient boosting algorithm that builds models sequentially to minimize errors. It incorporates regularization to prevent overfitting while improving efficiency and robustness. XGBoost is highly versatile, supporting multiple loss functions and handling missing data, making it ideal for a wide range of machine learning challenges.
\end{enumerate}
We further enhance predictive performance by combining these three classifiers through ensemble learning. Meanwhile, to obtain the likelihood value for each node pair, we perform 5-fold cross-validation to arrive at a final likelihood estimate (Fig.~\ref{fig2}(f)).

{\bf Constructing a Weighted Similarity Network}. The likelihood values derived from the machine learning models are integrated back into the original network. Recognizing the bidirectional interactions between node pairs, we square each likelihood value to calculate the similarity between nodes. This generates a weighted similarity network (Fig.~\ref{fig2}(g)).

{\bf Community Detection and Evaluation}. We finally apply four community detection algorithms to the weighted similarity network and evaluate the effectiveness of our framework using three evaluation metrics (Fig.~\ref{fig2}(h)).

Our framework employs a statistical-physics detection algorithm to infer community structures in the absence of ground-truth data, providing both theoretical and practical insights into the relationship between mesoscopic network properties and microscopic node pair similarities. By integrating mesoscopic structural information with microscopic feature-based similarities through machine learning, our approach offers an efficient and accurate method for community detection, yielding improved results compared to existing methods.

\begin{table}
\centering
\caption{Statistics of datasets. $|N|$ and $|L|$ denote the number of nodes and links, respectively, while $k$ represents the number of communities in the network. $CC$ refers to the average clustering coefficient of the network.}
\label{tab:table1}
\begin{tabular}{cccccc}
\toprule
Datasets & $|N|$ & $|L|$ & $k$ & $CC$ \\
\hline
Tree &100& 200 & 7 & 0.003 \\
LOL  & 55 & 174 & 10 & 0.745 \\
Email  & 1,005 & 25,571 & 42 & 0.399\\
YouTube & 49,236 & 298,335 & 6,221 & 0.177 \\
DBLP & 259,065 & 944,039 & 9,448 & 0.646 \\
Amazon & 316,164 & 868,929 & 2,572 & 0.399 \\
\bottomrule
\end{tabular}
\end{table}

\section{Experiments and analysis}
\label{sec:level4}

\subsection{Datasets}
\label{section41}

\subsubsection{Artificial networks}
We test our framework on a tree-like benchmark network which produces networks with properties of real-world networks \cite{shang2022link}. Here, we generate a tree-like network with well-defined community structures based on the following parameters:
\begin{itemize}
    \item $m=1$: Each newly added node forms one new link.

    \item $n=100$: The total number of links in the network is 100.
    
    \item $\beta$: A tunable parameter that controls the probability of forming a new community at each step. This probability is equal to the likelihood that a new node joins the smallest existing community, calculated as the size of the smallest community divided by the total number of nodes.
    
    \item $p=0.9$: Control the proportion of intra-community and inter-community links, with 90\% of links formed within communities and 10\% between communities.

    \item $k=3$: The initial network consists of three nodes forming a triangular structure.
\end{itemize}

\subsubsection{Real-world networks}
We conduct community detection experiments on four widely used large-scale networks with ground-truth communities from SNAP \cite{jure2014snap}. The Email dataset is derived from email records of a large European research institution \cite{yin2017local}. In this network, nodes represent institution members, and links indicate that one individual has sent at least one email to another. Ground-truth communities correspond to the institution’s 42 departments, with each member belonging to exactly one. In the Amazon network, nodes represent products, and links denote frequent co-purchases \cite{yang2012defining}. Ground-truth communities are based on product categories provided by Amazon, with each connected component within a category treated as a separate community. The DBLP dataset constructs a co-authorship network where two authors are connected if they have co-authored at least one paper \cite{yang2012defining}. Ground-truth communities correspond to publication venues, such as journals or conferences, with each connected component within a venue treated as a separate community. The YouTube social network consists of users who form friendships and create groups that others can join \cite{yang2012defining}. Ground-truth communities are defined by these user-created groups, with each connected component within a group considered a distinct community.

The four datasets are undirected and unweighted networks. The Email dataset provides non-overlapping community information for the four datasets, whereas the Amazon, DBLP, and YouTube datasets contain overlapping communities. Since this study focuses on non-overlapping community detection, we preprocess the original data to transform overlapping communities into a non-overlapping structure. The preprocessing procedure consists of the following steps:
\begin{enumerate}


    \item Assign each node to its largest associated community. If a node belongs to multiple communities of the same maximum size, one is selected at random.

    \item Calculate the current community size. If a community consists of only one node, it is removed.

    \item For the link list, retain a link only if both of its nodes belong to communities with more than one member; otherwise, remove the link.

\end{enumerate}

In addition to the four publicly available real-world datasets, we construct the LOL dataset using the rosters of 10 teams from the League of Legends Champions Korea between 2022 and 2024. In this dataset, nodes represent players, and links indicate teammate relationships formed during this period. The team composition in 2024 serves as the ground-truth community structure. The statistics of all datasets we deal with are shown in Table~\ref{tab:table1}.

\begin{table*}
\centering
\caption{Performance comparison of the average $Q^w$ across different methods for community detection. The best results from the comparison between the original detection method and the statistical-physics approach are highlighted in bold. Here, we report the best detection performance among 16 results of each network.}
\label{tab:qw}
\begin{tabular}{cccccccc}
\toprule
 Rule&Method&Tree&LOL&Email&YouTube&DBLP&Amazon\\
\hline
\multicolumn{1}{c}{\multirow{4}*{Original}}& Louvain & 0.795 & 0.673 &0.413& 0.628 & 0.810 & 0.928 \\
& Infomap & 0.707 & 0.673 &0.399& 0.637 & 0.808 & 0.438 \\
& Leiden & 0.796 & 0.673 &0.416& 0.637 & 0.821 & 0.933 \\
& FG & 0.793 & 0.658 &0.341& 0.589 & 0.719 & 0.877 \\
\hline
\multicolumn{1}{c}{\multirow{3}*{Statistical physics}}& DT-Leiden\footnotemark[1] & 0.815 & {\bf0.749} &0.483& {\bf0.651} & 0.852 & {\bf0.938} \\
& RF-Leiden & {\bf0.819} & 0.733 &0.494& {\bf0.651} & {\bf0.853} & {\bf0.938} \\
& VC-Leiden\footnotemark[2] & 0.814 & 0.735 &{\bf0.694}& 0.648 & 0.847 & 0.937 \\
\bottomrule
\end{tabular}
\footnotetext[1]{Since the proposed detection framework integrates multiple algorithms, we designate specific methods based on the sequential steps of the framework. For instance, DT-Leiden signifies that the decision tree (DT) algorithm is employed in step Fig.~\ref{fig2}(e), while the Leiden algorithm is utilized for community detection in step Fig.~\ref{fig2}(g). In the statistical physics-based approach, the community detection algorithms remain consistent across steps Figs.~\ref{fig2}(b) and (g).}
\footnotetext[2]{To improve detection performance, we integrate Decision Tree (DT), Random Forests (RF), and XGBoost (XGB) using two common ensemble strategies: soft voting and hard voting. Here, VC denotes the optimal result obtained from either strategy. Hence, we can generate 16 results using the above naming convention.}
\end{table*}

\subsection{Results on statistical-physics approaches}
\label{section42}
From a theoretical perspective, modularity is a key metric for assessing the quality of community structures. A higher modularity score indicates that the detected communities have stronger internal connectivity while remaining well-separated from each other. The statistical-physics methods, particularly RF-Leiden and DT-Leiden, consistently achieve higher modularity scores across all datasets, indicating better community detection performance (Table \ref{tab:qw}). For example, RF-Leiden outperforms the best Original method (Leiden) by 0.023, achieving a modularity score of 0.819 compared to 0.796 on the Tree dataset. On Email, VC-Leiden reaches a modularity score of 0.694, which is a notable improvement over the best original method (Leiden), which scores 0.416.

The statistical-physics methods in our framework benefit from integrating ensemble methods such as Decision Trees (DT), Random Forests (RF), and XGBoost (XGB). By using strategies like soft and hard voting, the framework combines the strengths of different algorithms to improve accuracy and stability. This results in more reliable community detection outcomes, particularly evident in the high scores on datasets such as LOL, Email, and DBLP.

The statistical-physics approach outperforms the original methods in a more consistent manner. While traditional methods like FG and Infomap show fluctuations in performance depending on the dataset, the statistical-physics methods maintain superior performance across all datasets tested. This consistency reflects the robustness of the statistical-physics approaches in handling different types of network structures and community characteristics. This can be attributed to the framework's ability to incorporate both mesoscopic (community structure) and microscopic (node-pair similarity) information, further enhanced by the ensemble strategies.

The statistical-physics approach shows clear advantages over original methods in terms of both performance consistency and modularity scores. The integration of ensemble learning techniques plays a significant role in improving community detection accuracy. These findings suggest that the statistical-physics approach provides a more effective and reliable solution for community detection, outperforming traditional methods across a variety of datasets. This supports our hypothesis that node-pair similarity is an important factor in enhancing community detection performance.

\subsection{Results on ground-truth approaches}
\label{section43}

Network modularity provides an intuitive understanding of community structure, but it does not directly reflect how well the detected communities align with ground-truth labels. To assess the practical accuracy of our framework, we evaluate it using NMI and ARI, which measure the agreement between detected communities and real-world classifications. 

The ground-truth methods exhibit clear advantages over the original methods in terms of the NMI metric. Across various datasets, the ground-truth methods consistently achieve higher NMI scores, indicating better community detection performance (Table \ref{tab:nmi}). For example, the  RF-FG achieves an NMI score of 0.254, outperforming the best original method (Leiden) with an NMI of 0.217 on the Tree dataset. On DBLP, VC-FG achieves an NMI score of 0.640, compared to the original method Infomap's 0.410.

Similarly, our framework achieves the highest ARI score, indicating a strong correspondence between the detected and actual communities (Table \ref{tab:ari}). This suggests that the method effectively preserves the underlying organization of real networks. For example, RF-Infomap achieves an ARI score of 0.060, which is significantly higher than the best original method (Infomap) with an ARI of 0.020 on the Tree dataset. On Amazon, RF-Infomap achieves the highest ARI score of 0.342, compared to the best original method (Leiden) with an ARI of 0.143.

The improvements in NMI and ARI scores suggest that incorporating ground-truth data and node-pair similarity by machine learning leads to more accurate and reliable community detection. This highlights the importance of using advanced methods that leverage known community structures for achieving better community detection performance, especially when dealing with complex and large-scale networks.

\begin{table*}
\centering
\caption{Performance comparison of the average NMI across different methods for community detection. The best results from the comparison between the original detection method and the ground-truth approach are highlighted in bold. Here, we report the best detection performance among 16 results of each network.}
\label{tab:nmi}
\begin{tabular}{cccccccc}
\toprule
 Rule&Method&Tree&LOL&Email&YouTube&DBLP&Amazon\\
\hline
\multicolumn{1}{c}{\multirow{4}*{Original}}& Louvain & 0.215 & 0.907 &0.580& 0.485 & 0.351 & 0.486 \\
& Infomap & 0.181 & 0.907 &0.631& 0.292 & 0.410 & 0.300 \\
& Leiden & 0.217 & 0.907 &0.581& 0.490 & 0.366 & 0.502 \\
& FG & 0.209 & 0.837 &0.442& 0.466 & 0.323 & 0.421 \\
\hline
\multicolumn{1}{c}{\multirow{5}*{Ground truth}}& VC-Leiden & 0.246 & 0.898 &0.566& 0.598 & 0.617 & {\bf0.503} \\
& VC-FG & 0.242 & 0.889 &0.516& 0.582 & {\bf0.640} & 0.471 \\
& RF-Louvain & 0.254 & 0.885 &0.561& {\bf0.602} & 0.410 & 0.500 \\
& XGB-Infomap & 0.181 & {\bf0.954} &{\bf0.635}& 0.491 & 0.453 & 0.350 \\
& RF-FG & {\bf0.254} & 0.885 &0.513&0.582&0.394&0.474 \\
\bottomrule
\end{tabular}
\end{table*}

\begin{table*}
\centering
\caption{Performance comparison of the average ARI across different methods for community detection. The best results from the comparison between the original detection method and the ground-truth approach are highlighted in bold. Here, we report the best detection performance among 16 results of each network.}
\label{tab:ari}
\begin{tabular}{cccccccc}
\toprule
 Rule&Method&Tree&LOL&Email&YouTube&DBLP&Amazon\\
\hline
\multicolumn{1}{c}{\multirow{4}*{Original}}& Louvain & 0.017 & 0.731 &0.313&0.051&0.035&0.142 \\
& Infomap & 0.020 & 0.731 &0.284&0.016&{\bf0.042}&0.083 \\
& Leiden & 0.016 & 0.731 &0.319&0.055&0.041&0.143 \\
& FG & 0.014 & 0.526 &0.150&0.040&0.013&0.117 \\
\hline
\multicolumn{1}{c}{\multirow{4}*{Ground truth}}&RF-Infomap & {\bf0.060} &0.784&0.335&0.037&0.001& {\bf0.342} \\
& XGB-FG &0.019&0.647&0.333&{\bf0.070} &0.029&0.146 \\
& DT-Leiden &0.023&0.667&{\bf0.391}&0.062&0.036&0.144 \\
& XGB-Infomap &0.020&{\bf0.883}&0.380&0.041&0.040&0.132 \\
\bottomrule
\end{tabular}
\end{table*}

\subsection{Main results and explanation}
\label{section44}
\begin{figure*}[htbp]
\centering
\includegraphics[width=\textwidth]{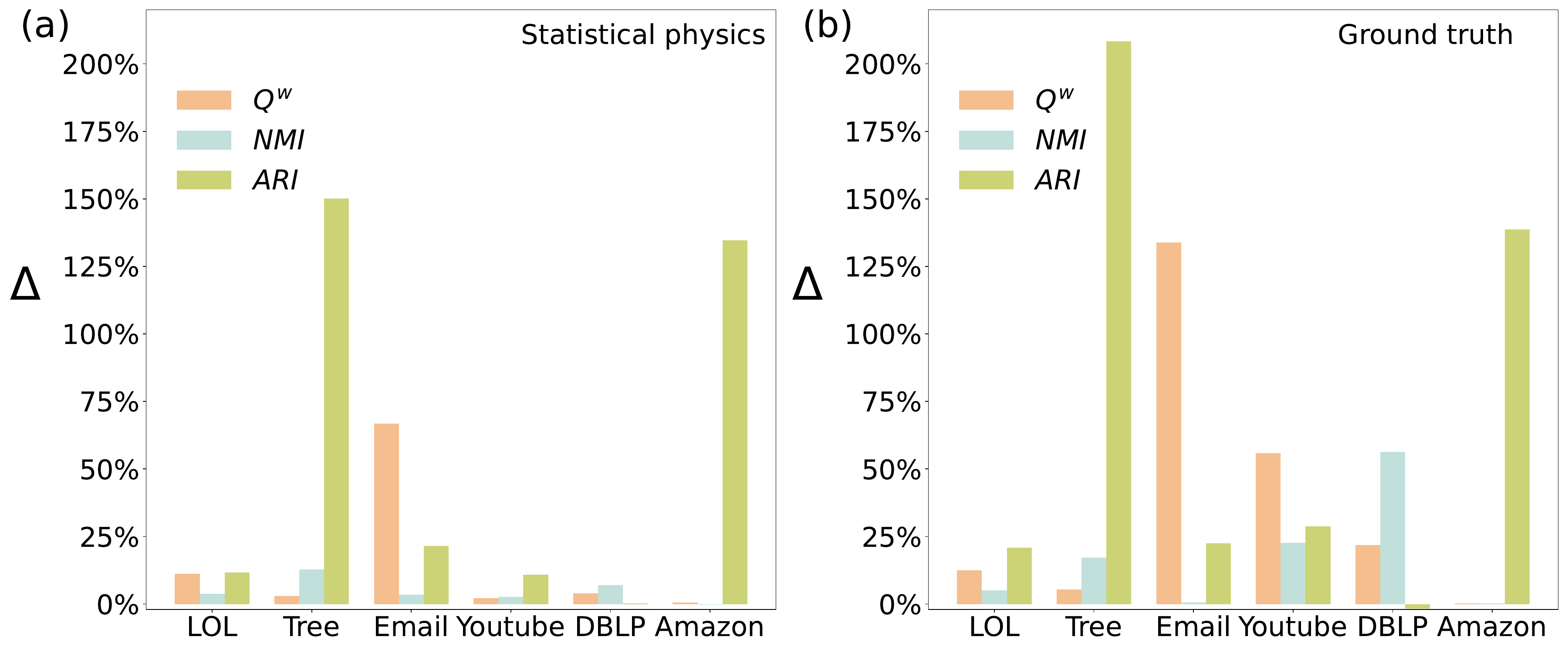}
\caption{Performance improvement of the proposed framework relative to the original method. Here, $\Delta$ represents the percentage improvement in detection performance under the statistical-physics or ground-truth approaches. It is calculated as the difference between the highest detection performance achieved by these approaches and that of the original method, normalized by the latter.}
\label{fig3}
\end{figure*}

Figure \ref{fig3} illustrates the percentage improvement ($\Delta$) in detection performance achieved by the proposed framework compared to the original methods across different datasets. Figure \ref{fig3}a presents the performance improvement under the statistical-physics approach, which is applied when ground-truth community information is unavailable. The results show a substantial increase in detection accuracy, particularly in ARI for the Tree and Amazon networks, where improvements exceed 125\%. Other networks also exhibit varying degrees of enhancement in $Q^w$, NMI, and ARI, demonstrating the effectiveness of the statistical-physics approach for community detection in the absence of predefined ground-truth communities.

Moreover, Figure \ref{fig3}b illustrates the performance improvement when ground-truth community information is available. The proposed ground-truth approach achieves significant gains across all three metrics, with ARI for the Tree dataset improving by nearly 200\%. Additionally, consistent enhancements in $Q^w$, NMI, and ARI across other networks further validate the benefits of incorporating ground-truth information when accessible.

Overall, Figure \ref{fig3} demonstrates that our framework consistently outperforms the original method, with the ground-truth approach yielding the most substantial improvements. These findings highlight the effectiveness of integrating machine learning to combine micro-level similarity information with mesoscopic community structures, leading to optimal detection performance. This underscores how machine learning can complement statistical physics, mitigating its limitations and facilitating globally optimal solutions.

To explain the performance improvement of our framework, we analyze the correlation between node-pair similarity and evaluation metrics (NMI and ARI). Specifically, we quantify this correlation using data from step (g) of our framework (Fig.~\ref{fig2}(g)), where node-pair similarity and evaluation metrics are derived from the best detection results. Table \ref{table:rela} reveals a strong correlation between node-pair similarity and evaluation metrics in both the statistical-physics and ground-truth approaches, with the correlation being particularly significant under the ground-truth approach. This finding highlights node-pair similarity as a key factor in enhancing community detection performance.

Although node-pair similarity and evaluation metrics exhibit a strong correlation under the statistical-physics approach, this correlation is not statistically significant. This is likely due to the assumption that initially detected communities serve as ground-truth benchmarks, leading to limited generalization and potential model instability. In contrast, under the ground-truth approach, the correlation is both strong and statistically significant, as actual community labels provide a reliable foundation, enhancing generalization. These findings suggest that when ground-truth communities are unavailable, our framework still surpasses traditional detection methods. More importantly, when ground-truth communities are accessible, our framework achieves a substantial improvement in community detection performance.

\begin{table}[htbp]
	\centering
	\caption{The correlation between node-pair similarity and evaluation metrics. To achieve this correlation, we quantify the relationship between $S^{in}-S^{out}$ and evaluation metrics (NMI and ARI) on all datasets with ground-truth communities. $S^{in}$ represents the average similarities of internal links, which are normalized by the sum of similarities between all node pairs and the total number of internal links within the community. Similarly, $S^{out}$ represents the average similarities of external links, which are normalized by the sum of similarities between all node pairs and the total number of external links connecting the community to others. The NMI and ARI are real values of a network. The r is the Pearson correlation coefficient, and the P-value is from the Student's t test.} 
     \label{table:rela}
	\begin{tabular}{l|rrrr}
            \toprule
		\multicolumn{1}{c}{\multirow{2}*{Rule}}& \multicolumn{2}{c}{NMI}& \multicolumn{2}{c}{ARI}\\
		   \cline{2-3} \cline{4-5}
		  \multicolumn{1}{c}{}& \multicolumn{1}{c}{r} & \multicolumn{1}{c}{P-value} & \multicolumn{1}{c}{r}& \multicolumn{1}{c}{P-value}\\
		  \hline		
		\multicolumn{1}{c}{\multirow{1}*{Statistical physics}} & \multicolumn{1}{c}{0.733}& \multicolumn{1}{c}{0.159} & \multicolumn{1}{c}{0.659}& \multicolumn{1}{c}{0.226} \\
        \multicolumn{1}{c}{\multirow{1}*{Ground truth}} & \multicolumn{1}{c}{0.827}& \multicolumn{1}{c}{0.084} & \multicolumn{1}{c}{0.876}& \multicolumn{1}{c}{0.052} \\
        \bottomrule
	\end{tabular}
\end{table}

\section{Conclusion and Discussion}
\label{sec:level5}
To summarize, we find that node-pair similarity and network modularity exhibit a strong and statistically significant correlation. Hence, we present a novel framework that integrates machine learning techniques to enhance community detection by incorporating micro-level node-pair similarities into mesoscopic community structures. Our approach bridges the gap between local node interactions and global network organization, leading to more accurate and robust community detection. By leveraging advanced ensemble learning models, we effectively embed node-pair similarity information into the detection process, improving the capture of nuanced structural dependencies.

Experimental results demonstrate the effectiveness of the proposed framework across multiple artificial and real-world networks. Theoretically, our method achieves the highest modularity scores, underscoring its ability to enhance the structural coherence of detected communities. Practically, when evaluated using accuracy metrics such as NMI and ARI, our approach consistently outperforms the original method, with particularly significant improvements under the ground-truth approaches. Specifically, we observe substantial performance gains across multiple datasets, indicating that our framework accurately recovers real-world community structures while minimizing incorrect assignments.

Our findings highlight the synergy between machine learning and statistical physics in community detection. While statistical physics provides powerful tools for modeling network structures, it has inherent limitations in capturing fine-grained node similarities. By integrating machine learning, we mitigate these limitations and achieve more optimal solutions. This integration is particularly beneficial when network community labels are available, enabling the most accurate detection results. 

Moreover, the teacher-student paradigm is an extremely popular approach for improving model performance in the field of computer science. Our research may have uncovered an evolutionary concept for this paradigm, where machine learning (the student) learns patterns from statistical physics methods (the teacher), and subsequently, the statistical physics models refine and enhance their understanding by learning from the outcomes produced by machine learning. This subsequent refinement suggests a cycle where the teacher (statistical physics methods) and the student (machine learning) can iteratively improve through mutual learning, indicating the feasibility of a teacher-student-teacher framework.

Despite its advantages, our approach has certain limitations. The performance of the framework may depend on the quality of node similarity features, as suboptimal feature selection could affect detection accuracy \cite{ran2024maximum}. Additionally, while validated on diverse datasets, its generalizability to networks with highly dynamic structures or varying connectivity patterns remains an open question \cite{holme2012temporal,li2017fundamental}. Future research could explore adaptive methods that dynamically adjust feature selection and learning strategies based on network characteristics. Furthermore, extending the framework to handle temporal or evolving networks would enhance its applicability in real-world scenarios.

\section{Acknowledgements}
This work is supported by the National Natural Science Foundation of China (62403062, 61803046), the Social Sciences Fund of Jiangsu Province 24XWB004, the Postdoctoral Fellowship Program of CPSF GZC20230281, the Fundamental Research Funds for the Central Universities 123330009, Ke-ke Shang is supported by Jiangsu Qing Lan Project.

\section{Data Availability Statement}
The original data is publicly available. The cleaned dataset and reproducible code will be made available to the public after the paper is accepted.


\bibliographystyle{plain}
\bibliography{main}

\end{document}